\documentclass[conference, letterpaper]{IEEEtran}
\usepackage{booktabs}
\usepackage{caption}
\ifCLASSINFOpdf
\else
\fi
\ifCLASSINFOpdf
\usepackage[pdftex]{graphicx}
\else
\fi

\usepackage[cmex10]{amsmath}
\usepackage{color}
\usepackage{fancyhdr}
\usepackage[caption=false,font=footnotesize]{subfig}

\renewcommand{\thispagestyle}[2]{}

\fancypagestyle{plain}{
	\fancyhead{}
	\fancyhead[C]{first page center header}
	\fancyfoot{}
	\fancyfoot[C]{first page center footer}
}
\begin{document}
\title{Towards Secure IoT Communication with Smart Contracts in a Blockchain Infrastructure}
\author{\IEEEauthorblockN{Jawad Ali}
\IEEEauthorblockA{Malaysian Insitute of \\ Information Technology, \\Universiti Kuala Lumpur, \\
Malaysia\\
jawad.ali@s.unikl.edu.my}
\and
\IEEEauthorblockN{Toqeer Ali}
\IEEEauthorblockA{Faculty of Computer \\ \& Information System \\ Islamic University of Madinah \\ toqeer@iu.edu.sa}
\and
\IEEEauthorblockN{Shahrulniza Musa}
\IEEEauthorblockA{Malaysian Insitute of \\ Information Technology, \\Universiti Kuala Lumpur,\\Kuala Lumpur\\
	Malaysia\\
	shahrulniza@unikl.edu.my}
\and
\IEEEauthorblockN{Ali Zahrani}
\IEEEauthorblockA{Faculty of Computer \\ \& Information System \\ Islamic University of Madinah \\ alzahrani@iu.edu.sa}
}
\maketitle

\begin{abstract} 
The Internet of Things (IoT) is undergoing rapid growth in the IT industry, but, it continues to be associated with several security and privacy concerns as a result of its massive scale, decentralised topology, and resource-constrained devices. Blockchain (BC), a distributed ledger technology used in cryptocurrency has attracted significant attention in the realm of IoT security and privacy. However, adopting BC to IoT is not straightforward in most cases, due to overheads and delays caused by BC operations. In this paper, we apply a BC technology known as Hyperledgder Fabric, to an IoT network. This technology introduces an execute-order technique for transactions that separates the transaction execution from consensus, resulting in increased efficiency. 
We demonstrate that our proposed IoT-BC architecture is sufficiently secure with regard to fundamental security goals i.e., confidentiality, integrity, and availability. Finally, the simulation results are highlighted that shows the performance overheads associated with our approach are as minimal as those associated with the Hyperledger Fabric framework and negligible in terms of security and privacy.

\end{abstract}

\begin{IEEEkeywords}
IoT, Blockchain Authorization, Hyperledger Fabric, BC, Blockchain Integrity.
\end{IEEEkeywords}

\section{Introduction}
The Internet of Things is undergoing exponential growth as everything is increasingly connected via the internet. According to the Gartner research report that predicts the future of IoT, there were almost 7 billion devices connected via smart technology in 2017, and this is set to approach 20 billion by 2020 \cite{gartner}. 
Some areas that apply this technology in daily life include automatic vehicles, home appliances, smart grid stations, health care applications, the retail sector, industrial supply chains and logistics management, security and surveillance, transportation and general industrial control systems  \cite{internet-app} \cite{internet-app2}. These all comprise a smart infrastructure supported by heterogeneous entities including web servers, end users, smartphones and cloud resources, that function with and without human interaction, selecting and providing information to the end users.

Millions of sensors and actuators are used in smart-environments to control and monitor lighting and heating systems, elevators, and cameras. Apart from this, Industrial control systems composed of numerous sub-systems, such as Supervisory control and data acquisition (SCADA), distributed control systems (DCS), other smaller systems, including programmable logic controllers (PLCs), remote terminal units (RTUs), and others involved in running an industrial operation \cite{indus}. These systems are composed of various interconnected sensing devices, and their heterogeneous nature and swiftly developing technology results in security and privacy risks that pose a major challenge to the IoT community.

With the widespread adoption and advancement of this technology, IoT devices face numerous security problems in terms of hardware, software and network communications. Increased communication between IoT devices involves large amounts of critical data and privacy-sensitive information that are increasingly vulnerable: a recent DDoS attack known as \emph{mirai} \cite{mirai} affected millions of IoT devices. Several approaches have been proposed to optimise security and privacy \cite{continous2015}\cite{Daojing2010} \cite{11sivaraman2015network}, but due to the IoT’s rapid expansion at a massive scale, no consensus has been reached regarding the optimal solution, and several questions remain.
 
In IoT systems, the majority of communication between devices is facilitated through client/server architecture or centralised architecture, through which authentication and identification, and other security measures, are processed by a central authority. The first major issue in these centralised systems, where everyone is reliant on a central authority, is the possibility of a single point of failure. Second, all connected devices must communicate via the internet regardless of the distance between them, thus leading to command processing overhead. Additionally, the current approach to IoT security involves high maintenance costs in terms of central cloud servers and other network equipment. 
In short, to migrate the current IoT centralised architecture to a decentralised approach some fundamental capabilities are required:
\begin{itemize}
	\item peer-to-peer communication,
	\item distributed file sharing
	\item autonomous device communication, and
	\item efficiency and security.
	
\end{itemize}
The above requirements are conveniently available in a newly introduced Blockchain (BC) technology known as Hyperledger Fabric \cite{hyperfabric}. BC is a shared and replicated ledger that offers immutability, consensus, and finality. Although there are several other BC technologies, BC such as that adopted by Bitcoin, but Fabric reduces computation cycles and provides scalability, identity management, and privacy, which is of the utmost importance in the IoT paradigm. The cryptographic algorithms used in BC ensure superior security and privacy for device data and IoT is increasingly adopting BC technology. As far as BC’s applicability to IoT is concerned, it records the transaction histories of smart devices in an immutable manner, thus eliminating the need for a central authority. For messaging exchange between IoT-devices a smart contract is used for successful agreement between all parties. The most promising feature of BC, with regard to its application in IoT, is the possibility of maintaining a transaction record across all devices, thus and thus make a distributed and trustless ledger.

Generally, all currently adopted approaches are highly centralised and thus unable to manage the IoT environment at the current scale. BC technology has the potential to overcome the scalability and security issues associated with IoT through decentralisation.
This paper’s contribution is twofold: To implement Hyperledger Fabric in a smart-IoT environment to assess the validity of the communicating devices whether normal or malicious i.e., to assure users of the integrity of the data from a particular device. Another important issue is that the IoT network is growing very rapidly as predicted, and cannot, as such, be efficiently and securely managed by the current centralised mechanisms. In this study, we implemented IoT-based architecture in tandem with BC (Hyperledger Fabric). We tested our scheme in a smart home-based scenario, however, it will be applicable in other contexts, including smart cities and smart industries. 

The remainder of this paper is organized as follows: Section II presents some background information concerning BC, smart contracts and Hyperledger Fabric. In section III, we detail the need for security in IoT with a review of the literature. Section IV presents our proposed architecture in detail. Section V includes the experiment results and some discussion of the findings. Finally, in section VI we conclude and propose potential future directions.

\section{Background}
\subsection{Block-Chain}
Blockchain (BC) is initially adopted by a very well-known crypto-currencies called bitcoin \cite{bitcoin}. BC record the transactions across many computers and store in a decentralised way and thus form a immutable digital distributed ledger. This ledger is distributed among all the nodes in the network and every one have the copy of all the transactions record. Blockchain is composed of two kinds of record: blocks \& transactions. Every transaction is recorded in the form of blocks and these blocks are then organized into a linear sequences and form a merkle tree. Every block in BC contains a hash of the previous block. New transactions are initially process by \emph{miners} which are further append to end of the chain and cannot be modified or remove by anyone, once accepted. Initially the BC is started with the first block called genesis or initial block. After then each transaction is validated stored in the form of block. Every block contains the hash to the preceding block and changes to the previous block would produced different hash-code and thus visible to all member in BC immediately. Therefore, the block-chains are considered to be tamper-proof ledger. Figure 1 shows a sample block chain.
\begin{figure}[!t]
	\centering
	\includegraphics[width=80mm]{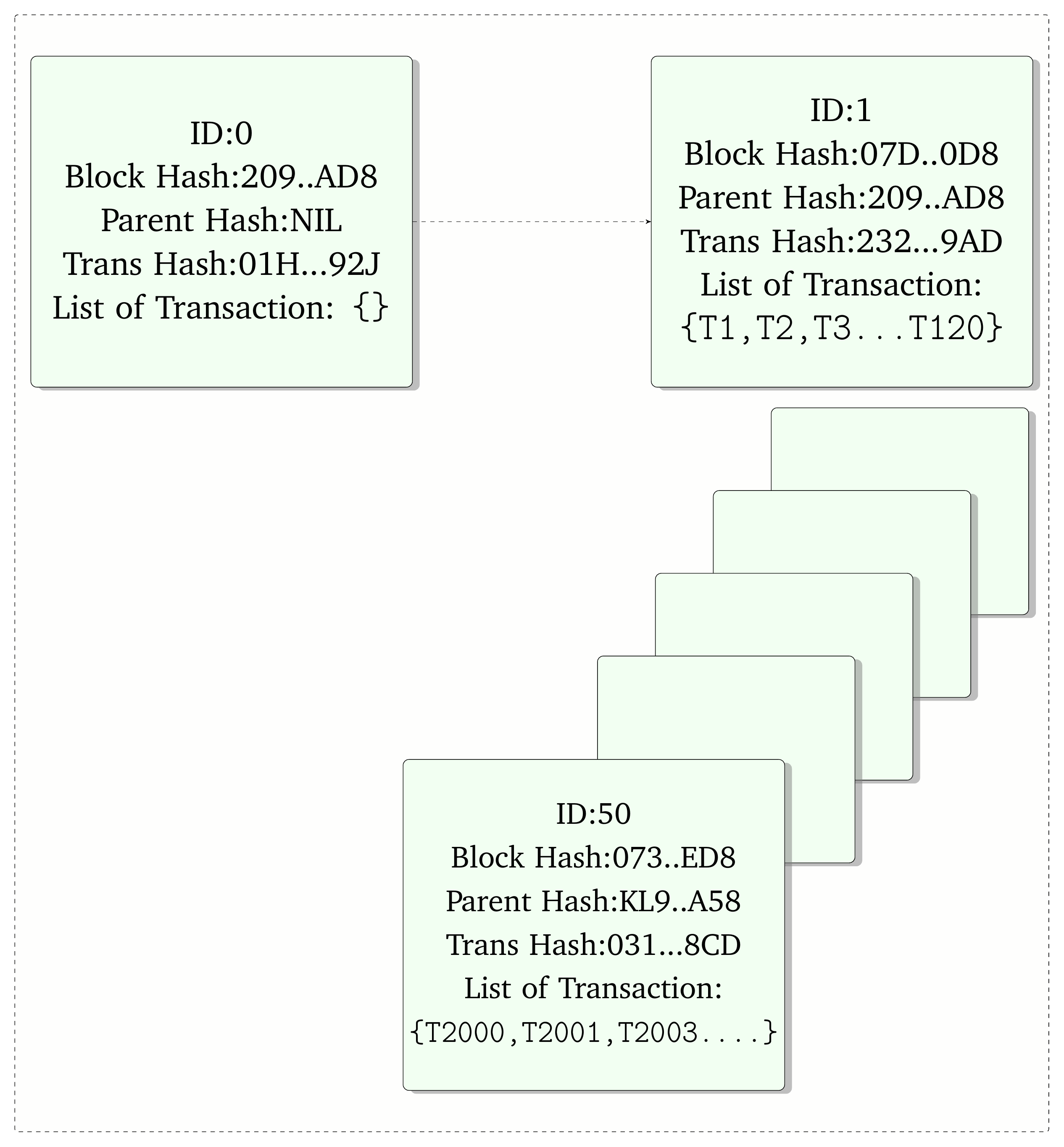}
	\caption{A sample Blockchain}
	\label{fig:bc}
\end{figure}

\subsubsection{BlockChain and IOT}
Authors in \cite{bciot} put efforts on the case of expanding IoT-devices to a decentralised way in order to be sustainable. From manufacturers perspective, current centralised mechanisms needed a high maintenance cost in case of distribution of software updates to a billion number of nodes in IOT. On the consumer's end, trust on devices is merely a big challenge where someone needs a transparent security. In short, these issues cannot be tackle without a trust-less and peer-to-peer architecture that can distribute data in a secure and transparent way. In \cite{Christidis2016}, authors argue that in current arising security problems with IOT, BlockChain provides a better an elegant solution.
\subsection{Smart Contracts} 
Smart Contracts \cite{smart-contract} also called crypto-contract are a computer program that are used to transfer/monitor assets or digital currencies among parties under certain rules. It does not only determine the conditions and penalties but can also enforce those policies/agreements. These smart contracts are stored on block-chains and BC is an ideal technology to store these contracts because of its immutability and security. Whenever a transaction is suppose to happen, smart-contract determine where the transaction should be transfer/returned or from where the transaction was actually originated.
\subsection{Hyper-Ledger Fabric}
Hyper-ledger fabric \cite{hyperfabric} is recently proposed by IBM team and considers it as an open-source blockchain platform and a truly scalable system for distributed applications running on a large-scale network. The main reasons behind this proposal is to cope with the limitations found in the other permissioned BlockChain architectures. Unlike all the other BC technologies that rely on order-execute paradigm while doing some kind of transactions, \emph{fabric} introduces execute-order paradigm where transactions executed at the first stage before reaching to their final order. Some of limitations that fabric overcome are as follows:
\begin{itemize}
	\item Performance overhead: The order-execute method is widely acceptable because of its simplicity, but it execute transactions sequentially on all peers in the network which may cause a performance overhead. A smart-contract with infinite loop may launch a DoS attack, which can severely degrade the performance on blockchain. This issue is resolve with fabric by execute the transaction first and then order the transaction over the BC. One of important benefit of execution at first is to be efficient and avoid DOS attack by denying the unvalidated transactions before ordering.
	\item Confidentiality: As discussed earlier that the design of other BC technologies runs the smart contracts on all the peers. However, in many scenarios of permissioned BCs, requires \emph{confidentiality} i.e. to restrict the ledger state or transaction data. In Fabric the execution of smart-contract restricted to a number of trusted peers (endorsers) that give assurance to the outcomes of execution. 
\end{itemize}
The transactions flow in Fabric comprises three different phases: Execution, ordering and validation. 
\begin{itemize}
	\item Execution Phase: In this first phase, the node in BC signs and send the transaction-proposal for execution to endorser(s) (designated peers). The information of endorsers are defined in the endorsement policy already. The endorsers is suppose to simulate the transaction with respect to the operations specified on the chaincode. After this simulation, each endorsers generate two values i.e. \emph{writeset} value that contains the state-update information and \emph{readset} defining the version of transaction-proposal simulation. Finally, the endorser(s) signs and send the message now called \emph{endorsement(s)} that composed of readset and writeset, to the client. The client receive the message and wait until they satisfy the chain-code endorsement policy. In particular, all the endorser(s) need to produce the same execution result. After this, the client may proceed to make a transaction and pass it to the ordering phase.
	\item Ordering Phase: Upon adequate endorsement on transaction-proposal, client assembles the \emph{transaction} and send it to the ordering service. This transaction consists of all the parameters including chain-code information, transaction metadata and set of endorsements. The ordering service organize total order of transactions per channel (sub-network) that has been submitted. Furthermore, the ordering service combine multiple transaction and arrange it into \emph{blocks} and form a blocks that improves the overall throughput of broadcast protocol. As it is known that there are a huge numbers of peers in BC, only few of them are supposed to implement ordering service. Finally, the ordering service also include the access control policies to check the clients whether it can receive block or not. 
	\item Validation Phase: When the ordering service sends the block to the peer, it will enter to validation phase for further checking that is done by three sequential steps. (1) Endorsement policy evaluation occurs, if it is not satisfied then the transaction disregarded and marked it as invalid. (2) To check read-write conflict by comparing the key version to the current ledger state if it is not matched then marked invalid and disregarded. (3) Finally, the ledger update phase run and append the block to the block-chain. 
\end{itemize}

\section{Security in IoT}
In the past decade, Internet of Things provides an automation and smartness in our daily lives.
Due to the rapid advancement and deployment of IoT in the current infrastructure leads to many security and privacy requirements. S.Sicari et al. \cite{sicari} outline the current problems in the existing IoT and analyze to the most relevant solutions. The key security requirements are confidentiality, authentication and access control. The author discussed various scheme of confidentiality and authentication in context of Wireless sensors networks (WSN), but because of heterogeneity and low power consumption in IoT several questions arises that are:
\begin{itemize}
	\item Are the WSN easily adaptable to IoT with it heterogeneous nature of devices and different applications.
	\item Which communication layer is responsible for authentication process.
	\item How to ensure the integrity and privacy of end-to-end devices in order to prevent from malicious attacks. 
\end{itemize}
Furthermore, access control that deals with user-access policies and usage control in the wide area network such as RBAC, DAC, MAC and their extensions. In context of IoT we needed to deal with streaming of data, Unlike traditional DBMS, where we deal with the discrete form of data. The problem arises here is the that the access control mechanism on data stream need more computation and hence lead to performance issue. Several solutions related to access control with data stream are discussed by the author and found the main challenges lying in it. Some issues are:
\begin{itemize}
	\item How to ensure that the things could be recognized by the system where user is not available.
	\item How to manage with large volume of data stream in a standard recognized way.
	
\end{itemize}
\subsection{Privacy \& Trust in IoT}
Keeping in mind the versatility of IoT in various areas that are: Smart-Homes, Smart-Cities, Smart-Health care, Traffic Control, smart parking system and so on. For all these scenarios, everyone need their protections for his personal information related to their movement from one place to another and communication with different other peoples. Author in \cite{atzori2010internet} investigated a survey study on all approaches in context of privacy in IoT and find a wide gap of research that need to be addressed before deployment in the current IoT paradigm.
The next important factor of security is satisfaction of trust between parties while communication. This satisfaction is basically reflected to the issues of identity management and access control mechanisms. In addition, they put an overview of trust management in IoT and discuss some past researches related to it. But all of them lacks a fully dynamic and distributed approach that can be suitable for the upcoming scale of IoT. Furthermore, most important missing item and a great challenge is a well-definition of trust agreement which needs to be a common language.
\subsection{Literature study}
David Airehrour \cite{Airehrour2016} did a thorough study on the security and privacy while routing
in internet of things. On security in IOT they find out the privacy preservation as
a key issue because of heterogeneity in devices over the network. Due to limited
battery capacity in most of IOT devices, several devices are not capable to end with
the complete process of cryptography and authentication and argued for a robust
secure authentication system. In general they classified the threats into two main
groups:
\begin{enumerate}
	\item \emph{General Security threats in IoT network}: It includes the traditional threats present
	in network such as DDOS, Man-in-the-middle attack etc. However, considering
	the scalability and massiveness of IoT networks along with their complexities
	\& heterogeneity, these traditional system attacks become more bigger challenge
	in IOT environment.
	\item \emph{Threats specific to IoT security:} These threats are related to the interconnection
	between IOT devices. For example sensitive data \& record retrieving of a heart or brain patient, smart meter reading and forest-reading. Reading of such
	sensitive data may be compromised while communicating over the IoT network
	or may be results in malicious transmission to other nodes for misuse. 
	
\end{enumerate}
Furthermore, this survey study highlighted the routing protocols i.e. (6LowPAN
and RPL) and their lack-ness in regards of security and privacy. Finally, the author
deliver some key considerations for the researchers in order to design secure routing
protocols for IOT. Some of recommendations are: secure routing establishment, self-stabilization,
location privacy and so on. In conclusion this whole study has revealed
that the current routing mechanisms and its standards are insecure for IoT.

Yuichi Kawamoto et al. \cite{Kawamoto2015} puts his efforts on location based authentication scheme in
IoT. In this proposed architecture, they used ambient information for uniqueness
by collecting the data from IOT-nodes at certain place \& time. In his case the
ambient information doesn't bound to use the key-elements such as SSID (service set
identifier) and RSS (received signal strength) for the freshness of data. Such kind
of authentication methods are more useful in scenarios like confidential information
floating in military area or other secret meetings which is limited to some users. The
important factor in this work is the unique information collected from different nodes
that is further used to authenticate the devices over network. Thus, it is needed to
collect real-time data as much as possible from many points in order to accomplish
an accuracy in authentication. The two metric for authentication are as follows:
\begin{itemize}
\item \emph{Freshness of Data}: As it is known that the data and location of nodes is continuously
changes at interval of time, thus it is need to collect the real-time-data in
order to ensure the accuracy.
\item \emph{Density of data collection}: The amount of data collected from huge number of
point will lead to accuracy in the system. However, such gathering of information
from massive points could decrease the real-time performance of system.
\end{itemize}
To uncover the optimal solutions for accuracy and efficiency, it is needed to evaluate both these metrics for different occasions. The main limitations and open research
holes of this work are: (a) Real-time performance degradation in case of data collection for huge number
of nodes. (b) More characteristics features added to ambient information could resolve the
real-time collection as well as accuracy.

Qinlong et al. \cite{qinlong2017} addressing the issues of security and privacy in Fog computing.
Basically, fog computing access the benefits of both the cloud and IoT paradigms.
Similarly the security concerns (confidentiality \& access control) with this kind of
computing is also similar to Cloud and IoT. This proposed work is based on
ABE-attribute based encryption and ABS (attribute based signature) which are the
types of cryptographic techniques. ABE process features an access control policies
over vast majority of attributes for user to perform decryption that can
provide confidentiality and access control of data.

Current ABE-based solutions cannot have the capacity to authenticate custom (some)
users to update the encrypted data. Thus, to update the cipher-data one should prove
his validity by public key management to cloud service provider (CSP). However, it
would create much burden on CSP in order to maintain a key-list for identification.
To cope with this issue, ABS (attribute based signature) is used to provide help to CSP
in regards of user validity. For the above reasons, the authors uses the combination of
ABE \& ABS for fine-grained access control in Fog computing paradigm. In the first
step, user data is encrypted with access and update policies and then fused the data
to cloud servers via fog nodes. Afterwards, those users whose attributes satisfy the
required access policies are capable to decrypt the encrypted data. For this purpose,
CSP is designated to check the signature in order to ensure the data integrity by
verifying the user policies. The contribution of this research in terms of security analysis are to ensure Data confidentiality, to provide authentication, fine-grained access control mechanism and
collusion resistance.

Michael et al. \cite{plati} proposed a platform for transactive IoT blockchain applications with repeatable testing (PlaTIBART) that combines actor patterns with custom DSL (domain specific language) and test network management tool.
DSL defines the roles that different clients in our network have, based on the actor pattern. The advantage of DSL model is to implement a correct-by-construction design means, that allows the verification stage on the model, to check the internal consistency before any deployment is attempted e.g to prevent inconsistencies: two clients requesting the same port on the same host.

PLaTIBART uses Ethereum as its BC implementation, i.e DSL has Ethereum-specific required setting such as ChainID and GasLimit etc. Future implementation will be refactor these requirements on other BC platforms, i.e. Hyperledger, formal Verification of internal-consistency of a configuration file and a means of defining incremental adjustments to a test network by DSL.

IoT in context of security and privacy is a very challenging because of its low constraints and resources capabilities of the heterogeneity among them. Moreover, these devices retrieve, store and share huge amounts of data from our personal (smart-home) to industry level, and thus attracting the community to a significant privacy concern. In \cite{1arabo2012privacy} the authors declares different zones for privacy in order to classify diverse kinds of data. Each of the zone is associated with context based policy checking technique, that is checked by Home-Security-Hub before accepting to join or re-join data from device. However, they do not consider the case of bypassing the hub and access the device directly.

A research efforts proposed in \cite{openpds} that relies of safe answers or aggregations of data in which the user can send only a small data as much possible to the third party or service provider. For ensuring privacy their technique add noise to the data in the smart-home environment, but this noisy data could lead to inaccurate and harmful services. 

A study in \cite{15smart} have been done regarding that the house-hold devices are more prone to attacks by users devices such as smartphones. As per perception that the router or gateway should offer security perimeter that can prevent from internet attacks. Authors shows a demonstration on attacks that penetrate the smart home network by a smart-phone applications. Such kind of attacks could be able to modifies firewall and allow the external user to directly attack the smart-device. Thus, it is concluded that routers and firewall at the home-end are consider as poor protection against internet attacks and figure out the need of extra security features on IoT devices.

Currently the CSIRO team proposed a new approach of integrating Blockchain with IoT \cite{Dorri2016}. In his initial efforts they use smart-home technology in order to realize that how blockchain can be deployed to IoT. The blockchain is specifically use to provide an access control mechanisms of smart-devices transactions located at smart-home. This research provides some extra security features by introducing BC technology in IoT, however it lacks the concept of consensus algorithm that every mainstream BC technology must have. Moreover, this technique cannot provide a generalize form of Block-Chain solution to IoT use cases.

In the light of above discussion, there is no technique that leverage the standard Block Chain implementation for scale-IOT environment. We argue that this research is the first step towards a generalize BC solution for the emerging IoT area. For better understanding we consider smart-home at this time as a use-case. 
In the next section, we will discuss our proposed IoT-Fabric architecture. 

\section{Proposed Solution}
In the current literature of IoT security and privacy several kinds of security and privacy mechanisms have been proposed. All of these mechanisms rely on centralised approach so-called client/server paradigm. Using this approach, all the devices in IoT network are identified and authenticated through a central point or server. But this centralised way supports today small-scale IoT network and will not capable to respond in the growing IoT system in the near future. Due to this rapid growth in IoT, existing solutions are very expensive in terms of high maintenance cost and infrastructure related to central cloud servers and other networking equipments.

Decentralised mechanisms in IoT infrastructure would remove several issues discussed above i.e. allowing peer-to-peer communications between IoT nodes will significantly reduce the cost of installation and maintenance. It will also distribute the storage and computation load over the entire IoT network that can achieve efficiency and prevention from a single point of failure. We realize and implement Hyper-ledger Fabric: A Block-Chain technology in our proposed solution. The benefit of adopting BC in our research is three-fold: Decentralised or distributed, Permission-based and all-secure. In this paper, we use smart-phone case study in order to exemplify our proposed architecture. However it is pluggable and well-suited in other IoT case scenarios.

\subsection{Hyper-Ledger Fabric based IoT architecture}
As discussed above that we consider a smart-home based scenario for better understanding. In a typical smart-home, user is connected to certain number of IoT devices i.e. IP camera, thermo-state, smart-phone, smart-bulb and several other sensors.
The architecture shown in Figure \ref{fig:arch} includes the components namely the smart-home, Hyper-ledger fabric interface and Block-chain Peers / Orderers.
Smart-homes has equipped with a number of different smart-IoT devices. Every device can share, store and update their transaction data. The Fabric interface is used to provide distributed ledger where each transaction from IoT devices is store in the sequence and thus output in Blocks. Smart-contracts defines the policies where every transaction is checked against a set of pre-defined policies. Finally, the consensus algorithm make an arc over the entire transactional flow, that can provide the generation of agreements on the order and to verify the correctness of blocks.

\subsubsection{Initialization}
 Initialization of the network is the first step that needs before the transactions are made. All the smart-devices in smart-home need to store the device information in smart-contract. Information in smart-contract might consist of every device information and its endorsement policies according to our defined policy structure in application interface.
\subsubsection{Handling Transaction in Smart-home}
In a smart-IoT environment, communication between devices is either directly or with other external resources i.e. cloud. Each smart-device requests some other device that can serve some services e.g. if a smart-bulb needs data from the motion sensor to start automatically in case if someone get to entrance of home. For such direct communication between devices a shared-key is used, in order to achieve user control over every transactions inside smart-home. Smart-contract has list of all devices in a particular environment that can share data by using secure shared key.
\subsubsection{Transactional Flow in Fabric}
\begin{figure*}[!th] 
	\centering
	\includegraphics[width=170mm]{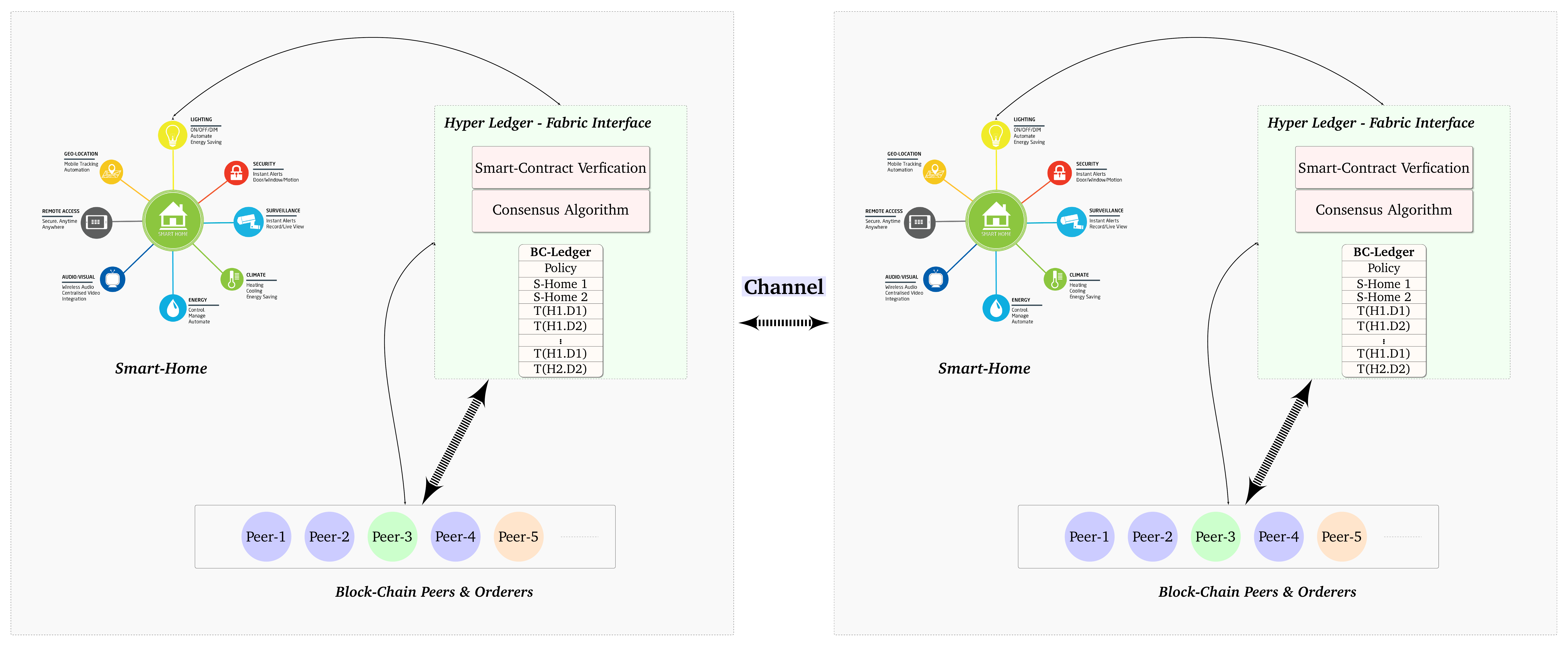}
	
	\caption{Proposed IoT Block Chain Architecture}

	\label{fig:arch}
\end{figure*}

A \emph{Hyperledger-fabric} consists of nodes or peers that constitute the network of Blockchain. As fabric architecture is permissioned, so every node or device participate using their identity provided by membership service provider (MSP). The flow of transactions in fabric are complete in three different phases i.e. execution, ordering and validation. In execution phase, a client (smart-home device) send a signed transaction proposal to one or more endorsers in the network. The endorsers (cf. Fig \ref{fig:arch})are specific peers that are defined in our smart-contract / chain-code via endorsement policy. After execution and get enough endorsements on proposal, client then assembles the transaction and submit to the ordering phase where all the transactions are properly placed in order. Moreover, the ordering phase also turned multiple transactions into blocks and also ensure that blocks on one channel are correctly ordered. Finally, the validation phase checks the validity of transaction sequentially as discussed in background section.

In our proposed architecture illustrated in Figure \ref{fig:arch}, we deploy smart-home based scenario on the Fabric architecture. The benefit of adopting Fabric is that, it totally separate consensus procedure from execution and validation which we need for efficient and scalable IoT environment. The transaction proposal begins from the smart-device to fabric interface where it first passed through our defined endorsement policy in smart-contract develop in Go language. The endorser peers recognizes and give decision on the transaction from a particular device and proceed for making transactions i.e store, access and monitor data. In the next step, as we discussed earlier that the ordering phase automatically broadcasts the endorsements in order to establish consensus and is responsible to make all the transactions in an orderly fashion and finally formed the shapes of Blocks. Finally validation phase runs and checks certain evaluation to prove that the transaction is valid or invalid.

Considers a smart-home based transaction (Figure \ref{fig:arch}) where a device send transaction data for storage on Blockchain e.g. \emph{store} transaction. The device first needs to prove his identity by matching the information defined in smart-contract. By receiving the device request or proposal the smart-contract identify the device and proceed for further procedure. After the device authenticated, it may send the data i.e. temperature data, along with the previous parameters (ID) for placement the transactions in order and store them to hash chain-sequence of block. The other possible types of transactions are \emph{monitor} and \emph{access} transactions. Such transactions are commonly initiated by user or simply smart-home owner in this case where he is far and need to access / monitor some data. 

Furthermore, \emph{channel} as depicted in \ref{fig:arch} is used to connect two different zones i.e more than one smart-home. However, at this time we are focusing on single home and multiple case scenario are out of scope in this research. In addition, we integrate our proposed architecture with an open-source distributed ledger known as IOTA \cite{iota}, that is specifically designed to power the upcoming future of IoT. It is known that per-missioned BC's like the one we used as a private BC for IoT network is although efficient but in a limited nodes. As the nodes increases in the IoT network the permissioned model of BC's degrades performance and hardly to scale-up in environment. As it is known that \emph{fabric} is pluggable architecture for in regards of consensus algorithms. For this purpose, the proposed architecture in this research is easy to integrate with DAG (Directed Acyclic Graph) for acquiring both side benefits i.e. Fabric for Efficiency \& DAG for scalability. DAG is recently adopted in IOTA a public BC having no mining fee, designed for IoT network. For more details regarding  IOTA \& DAG we refer the reader to \cite{tangle}. 

\section{Results and Discussion}
In this section, we detail our analysis on security, privacy and performance overhead in the smart-home use case. 
Before going to results and analysis section we discuss some reasons behind choosing Fabric in our proposed architecture.
Hyper-ledger Fabric proposed by IBM \cite{hyperfabric} has gain a lot of attention from market. Some of reasons behind choosing Fabric for our proposed IoT architecture are as follows:
\begin{enumerate}
	\item Generalized BC: As discuss earlier, an industry big players are supporting Fabric and make several solutions for it. Keeping this in mind, we argue that our proposed solution will be easily adaptable by other use cases in IoT and becomes a standardized Block-Chain technology for IoT-applications.
	\item Execute-order: Previous BC technologies relies on order-execute architecture where mining of transactions degrade performance. Fabric introduces execute-order where consensus phase is separated from execution, which ultimately improve performance. For IoT, performance is a key challenge and Fabric mitigate this issue comparing to other BCs.
\end{enumerate}

\subsection{Security Analysis}  
The three main requirement for robust security design in every case are mainly: Confidentiality, integrity and Availability or simply (CIA) \cite{cia}. Confidentially stated that only authorized entity must be granted access. Integrity makes sure that the data is not changed or modified at the receiving end and Availability means that the service or data to the user is available when someone has needed. All of these requirements are addressed in our proposed architecture. Table \ref{tab1} analyze that how our proposed solution could achieved the mainstream security requirements.
\begin{table}[h!]
	\centering
	\caption{Security Evaluation}
	\label{tab1}
	\begin{tabular}{@{}ll@{}}
		\toprule
		\textbf{Security Requirements} & \textbf{Solution Provided}    \\ \midrule
		Confidentiality               & Matching ID in Smart Contract \\ \midrule
		Integrity                      & Hashing Mechanism             \\ \midrule
		Authorization                  & Endorsement Policy checking   \\ \bottomrule
	\end{tabular}
\end{table}

Further, we need to analyze the famous attack known as DDOS (Distributed Denial of Service Attack) where an attacker uses several malicious or previous infected IoT-devices and attack some particular target device. A number of attacks have been discussed in \cite{cia} that can exploit the IoT-network in the form of DDOS attack. 
Our architecture is somehow provides a hierarchical prevention from such kinds of attacks. At the first level, attacker cannot gain access to these smart devices because these devices are generally not accessible physically. For instance, if attacker got access and supposed to infect the device, the second level of our security defense is that, each outgoing transaction has to be identified by our endorsement policy. Thus any transaction or traffic that constitute the DDOS attack would be rejected and cannot exit from the smart-home.
\subsection{Performance Evaluation} 
IoT with Block-Chain integration acquire extra computational overhead on the smart-devices in terms of packet communication, time overhead and Energy consumption. However, these low overhead somehow have no bigger impact and significantly increase security and privacy. For communication between devices IPV6 over (LoWPAN) is used in our smart-home devices because it is well-suited for low-resource devices in IoT. Secondly, Fabric is a complex architecture and the performance is depends on various parameters such as choice of distributed applications, transaction size, ordering services, consensus algorithm and network topology etc. Therefore, an in-depth performance analysis is not possible in this phase of research. This research is focusing on a standardize Block-Chain application for the growing need of IoT. 

However, the evaluation done in Fabric architecture \cite{hyperfabric} rely on crypto-currencies and is not directly applicable to our solution. To provide an evaluation for our use-case we simulate two type of transactions in a smart-home setting i.e. store and access. A store transaction means that a smart-device would like to store a transaction data on Fabric i.e. temperature sensor data, and access transaction is used to invoke some data.
In our simulation we analyze two types of experiments that are:
\subsubsection{Block Size Analysis}
\begin{figure}[h]    
	\centering   
	\fbox{\includegraphics[width=140px]{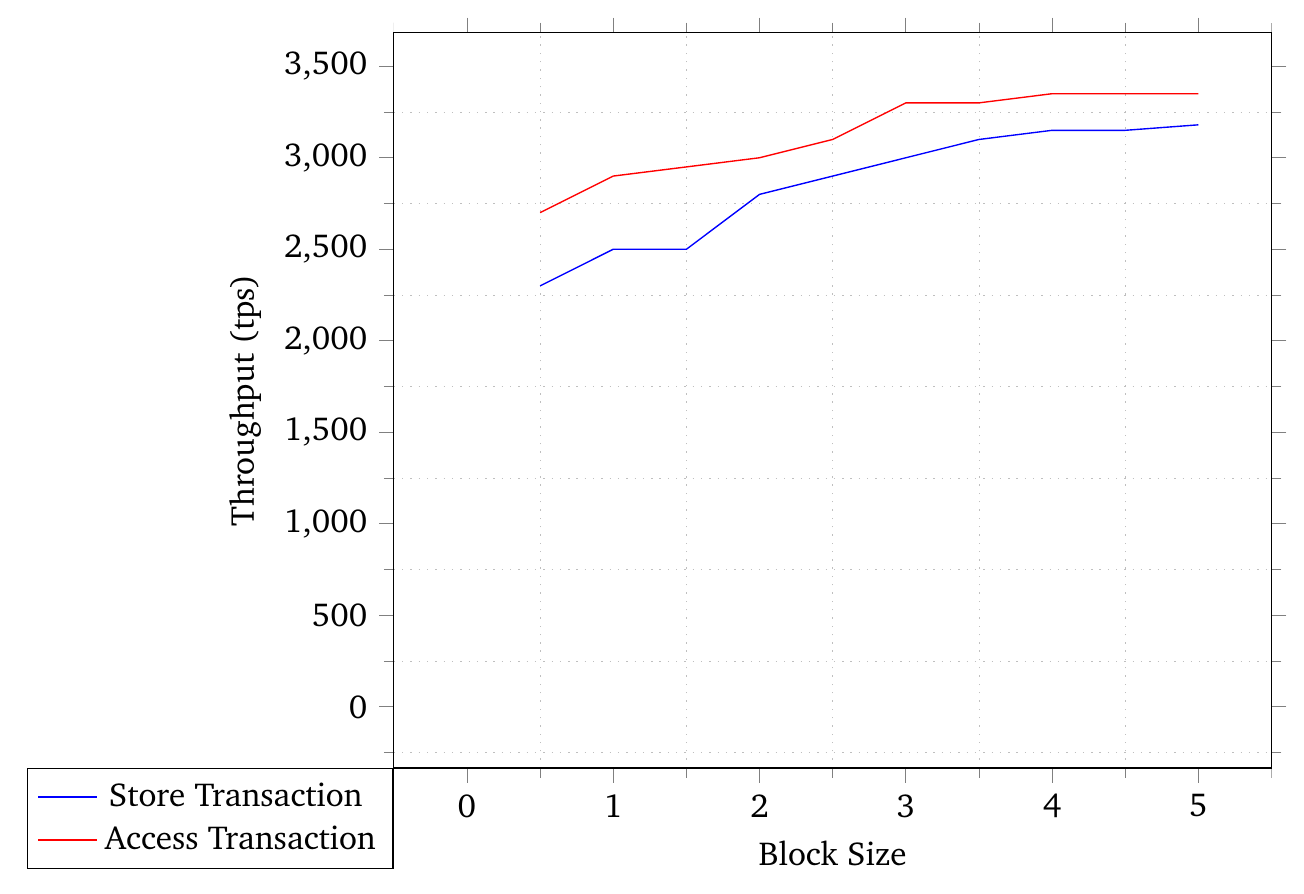}}   
	
\caption{Impact of Block Size on Throughput}
\label{throu}
	
	\fbox{\includegraphics[width=140px]{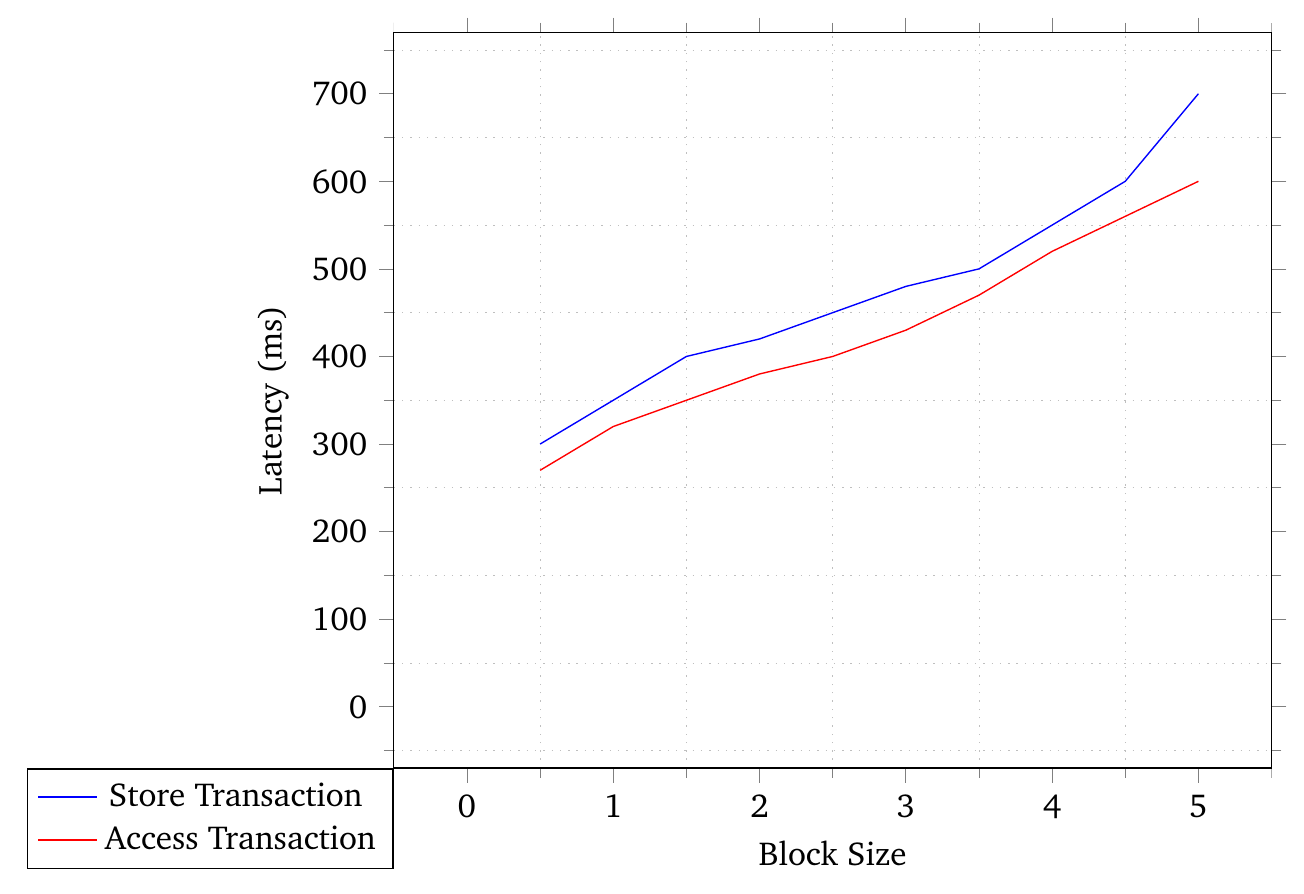}}
	
	\caption{Impact of Block Size on end-to-end Latency}
	\label{lat}
\end{figure}

Recall that transactions are combined and shaped into form of blocks. So the size of block is key factor that impact the overall throughput and latency. We ran simulation of block size varying 0.2MB to 4.5MB. Figure \ref{throu} show throughput measured at peers and Figure \ref{lat} illustrate end-to-end latency impacted by block sizes. We observe in our case that the block size of 2 to 2.5MB is significant in terms of throughput. However, the latency get worse as expected with increase in block size, but in this particular case we assume that roughly about 450 to 500 (ms) is acceptable. 

\subsubsection{Transaction Size}
We also investigate the number of transactions per block. Specifically, 1MB block contains 230 store and 350 access transactions. The sizes calculated are 3.00KB for store and 4.20KB for access transactions. In general, the transaction sizes in Fabric are larger because they also carry the certificate information for approval.
	
\subsubsection{Transaction Payload Size Analysis}
This experiment measures the latencies at different transactions payload sizes. Transaction payload is actually the size of data that is going to passed in a smart contract upon invoking. Table \ref{tab2} shows the comparative transaction payload analysis of Proposed IoT-BC and QUORAM-BC \cite{quoram}. QUORAM is also a permissioned public BC and based on Ethereum. As it is known that Ethereum \cite{ethereum} uses mining for consensus which is slow as compared to Fabric, and thus, not suitable for IOT environment. In QUORAM, they selected 32KB size for transaction payload, therefore we also considered the same size for comparison. The results show that the latencies increases with the increment of 10KB in payload size. The total increase in transaction latency in QUORAM 25.23\%, while in our technique the value is approximately 22.45\%. 

\begin{table}[]
\centering
	\caption{Latency with variable Transaction sizes}
	\label{tab2}
	\begin{tabular}{@{}clc@{}}
		\toprule
		\multicolumn{1}{l}{}                  & \multicolumn{2}{c}{\textbf{\begin{tabular}[c]{@{}c@{}}Latencies (s)\\ Transaction Payload\end{tabular}}} \\ \midrule
		\multicolumn{1}{l}{Payload Size (KB)} & Proposed BC                                & \multicolumn{1}{l}{Quoram BC \cite{quoram}}                               \\ \midrule
		1                                     & 0.225                                      & 0.325                                                       \\ \midrule
		10                                    & 0.280                                      & 0.383                                                       \\ \midrule
		20                                    & 0.320                                      & 0.384                                                       \\ \midrule
		30                                    & 0.330                                      & 0.407                                                       \\ \bottomrule
	\end{tabular}
\end{table}

\section{Conclusion and Future Work} 
Security and privacy in IoT is extremely important these days and gain a considerable attention from research and industry. Current security models for IoT are not suitable anymore due to rapid-scale, high maintenance cost of equipments, performance overhead and energy consumption. To cope with these problems, we proposed a new approach of implementing IoT Application on Fabric-BC.  Hyper-ledger Fabric introduces a novel framework in BC that separate the execution phase from consensus and implement policy-based endorsements. Our representative case-study through out research was smart-home. We presented solution for mainstream security requirements. We also discussed performance overhead of some transactions and found no extra overhead by our application interface developed on top of Fabric for IoT. Furthermore, a comparison has been done with QUORAM-BC which shows that our architecture is more efficient, specifically for IoT networks. Future work will include applications of our framework to other IoT domains, in-depth overhead analysis and integration with IOTA implementation.  

\bibliographystyle{IEEEtran}
\bibliography{jawadref}   

\begin{thebibliography}{10}
\providecommand{\url}[1]{#1}
\csname url@rmstyle\endcsname
\providecommand{\newblock}{\relax}
\providecommand{\bibinfo}[2]{#2}
\providecommand\BIBentrySTDinterwordspacing{\spaceskip=0pt\relax}
\providecommand\BIBentryALTinterwordstretchfactor{4}
\providecommand\BIBentryALTinterwordspacing{\spaceskip=\fontdimen2\font plus
\BIBentryALTinterwordstretchfactor\fontdimen3\font minus
  \fontdimen4\font\relax}
\providecommand\BIBforeignlanguage[2]{{%
\expandafter\ifx\csname l@#1\endcsname\relax
\typeout{** WARNING: IEEEtran.bst: No hyphenation pattern has been}%
\typeout{** loaded for the language `#1'. Using the pattern for}%
\typeout{** the default language instead.}%
\else
\language=\csname l@#1\endcsname
\fi
#2}}

\bibitem{gartner}
``{Internet of Things},'' \url{http://www.gartner.com/newsroom/id/3165317}.

\bibitem{internet-app}
N.~Shahid and S.~Aneja, ``Internet of things: Vision, application areas and
  research challenges,'' in \emph{I-SMAC (IoT in Social, Mobile, Analytics and
  Cloud)(I-SMAC), 2017 International Conference on}.\hskip 1em plus 0.5em minus
  0.4em\relax IEEE, 2017, pp. 583--587.

\bibitem{internet-app2}
I.~Lee and K.~Lee, ``The internet of things (iot): Applications, investments,
  and challenges for enterprises,'' \emph{Business Horizons}, vol.~58, no.~4,
  pp. 431--440, 2015.

\bibitem{indus}
\BIBentryALTinterwordspacing
K.~Stouffer, V.~Pillitteri, S.~Lightman, M.~Abrams, and A.~Hahn, ``{Guide to
  Industrial Control Systems (ICS) Security},'' 2015. [Online]. Available:
  \url{https://nvlpubs.nist.gov/nistpubs/SpecialPublications/NIST.SP.800-82r2.pdf}
\BIBentrySTDinterwordspacing

\bibitem{mirai}
``{Mirai Attack},''
  \url{https://www.corero.com/resources/ddos-attack-types/mirai-botnet-ddos-attack.html}.

\bibitem{continous2015}
\BIBentryALTinterwordspacing
O.~O. Bamasag and K.~Youcef-Toumi, ``{Towards Continuous Authentication in
  Internet of Things Based on Secret Sharing Scheme},'' \emph{Proceedings of
  the WESS'15: Workshop on Embedded Systems Security}, pp. 1:1----1:8, 2015.
  [Online]. Available: \url{http://doi.acm.org/10.1145/2818362.2818363}
\BIBentrySTDinterwordspacing

\bibitem{Daojing2010}
H.~Daojing, Y.~Gao, S.~Chan, C.~Chem, and J.~Bu, ``{An Enhanced Two-factor User
  Authentication Scheme in Wireless Sensor Networks},'' \emph{Ad Hoc {\&}
  Sensor Wireless Networks}, vol.~10, no.~4, pp. 361--371, 2010.

\bibitem{11sivaraman2015network}
V.~Sivaraman, H.~H. Gharakheili, A.~Vishwanath, R.~Boreli, and O.~Mehani,
  ``Network-level security and privacy control for smart-home iot devices,'' in
  \emph{Wireless and Mobile Computing, Networking and Communications (WiMob),
  2015 IEEE 11th International Conference on}.\hskip 1em plus 0.5em minus
  0.4em\relax IEEE, 2015, pp. 163--167.

\bibitem{hyperfabric}
\BIBentryALTinterwordspacing
E.~Androulaki, A.~Barger, V.~Bortnikov, C.~Cachin, K.~Christidis, A.~{De Caro},
  D.~Enyeart, C.~Ferris, G.~Laventman, Y.~Manevich, S.~Muralidharan, C.~Murthy,
  B.~Nguyen, M.~Sethi, G.~Singh, K.~Smith, A.~Sorniotti, C.~Stathakopoulou,
  M.~Vukoli{\'{c}}, S.~W. Cocco, and J.~Yellick, ``{Hyperledger Fabric: A
  Distributed Operating System for Permissioned Blockchains},'' 2018. [Online].
  Available:
  \url{http://arxiv.org/abs/1801.10228{\%}0Ahttp://dx.doi.org/10.1145/3190508.3190538}
\BIBentrySTDinterwordspacing

\bibitem{bitcoin}
S.~Nakamoto, ``Bitcoin: A peer-to-peer electronic cash system,'' 2008.

\bibitem{bciot}
P.~Brody and V.~Pureswaran, ``Device democracy: Saving the future of the
  internet of things,'' \emph{IBM, September}, 2014.

\bibitem{Christidis2016}
K.~Christidis and M.~Devetsikiotis, ``{Blockchains and Smart Contracts for the
  Internet of Things},'' \emph{IEEE Access}, vol.~4, pp. 2292--2303, 2016.

\bibitem{smart-contract}
``{Smart Contracts},''
  \url{http://searchcompliance.techtarget.com/definition/smart-contract}.

\bibitem{sicari}
D.~Miorandi, S.~Sicari, F.~De~Pellegrini, and I.~Chlamtac, ``Internet of
  things: Vision, applications and research challenges,'' \emph{Ad hoc
  networks}, vol.~10, no.~7, pp. 1497--1516, 2012.

\bibitem{atzori2010internet}
L.~Atzori, A.~Iera, and G.~Morabito, ``The internet of things: A survey,''
  \emph{Computer networks}, vol.~54, no.~15, pp. 2787--2805, 2010.

\bibitem{Airehrour2016}
D.~Airehrour, J.~Gutierrez, and S.~K. Ray, ``{Secure routing for internet of
  things: A survey},'' \emph{Journal of Network and Computer Applications},
  vol.~66, pp. 198--213, 2016.

\bibitem{Kawamoto2015}
Y.~Kawamoto, H.~Nishiyama, N.~Kato, Y.~Shimizu, A.~Takahara, and T.~Jiang,
  ``{Effectively collecting data for the location-based authentication in
  Internet of things},'' \emph{IEEE Systems Journal}, vol.~PP, no.~99, pp.
  1--9, 2015.

\bibitem{qinlong2017}
\BIBentryALTinterwordspacing
Q.~Huang, Y.~Yang, and L.~Wang, ``{Secure Data Access Control with Ciphertext
  Update and Computation Outsourcing in Fog Computing for Internet of
  Things},'' \emph{IEEE Access}, vol.~5, pp. 1--1, 2017. [Online]. Available:
  \url{http://ieeexplore.ieee.org/document/7981322/}
\BIBentrySTDinterwordspacing

\bibitem{plati}
\BIBentryALTinterwordspacing
M.~A. Walker, A.~Dubey, A.~Laszka, and D.~C. Schmidt, ``{PlaTIBART: a Platform
  for Transactive IoT Blockchain Applications with Repeatable Testing},'' 2017.
  [Online]. Available: \url{http://arxiv.org/abs/1709.09612}
\BIBentrySTDinterwordspacing

\bibitem{1arabo2012privacy}
A.~Arabo, I.~Brown, and F.~El-Moussa, ``Privacy in the age of mobility and
  smart devices in smart homes,'' in \emph{Privacy, Security, Risk and Trust
  (PASSAT), 2012 International Conference on and 2012 International Confernece
  on Social Computing (SocialCom)}.\hskip 1em plus 0.5em minus 0.4em\relax
  IEEE, 2012, pp. 819--826.

\bibitem{openpds}
Y.-A. de~Montjoye, E.~Shmueli, S.~S. Wang, and A.~S. Pentland, ``openpds:
  Protecting the privacy of metadata through safeanswers,'' \emph{PloS one},
  vol.~9, no.~7, p. e98790, 2014.

\bibitem{15smart}
V.~Sivaraman, D.~Chan, D.~Earl, and R.~Boreli, ``Smart-phones attacking
  smart-homes,'' in \emph{Proceedings of the 9th ACM Conference on Security \&
  Privacy in Wireless and Mobile Networks}.\hskip 1em plus 0.5em minus
  0.4em\relax ACM, 2016, pp. 195--200.

\bibitem{Dorri2016}
\BIBentryALTinterwordspacing
A.~Dorri, S.~S. Kanhere, and R.~Jurdak, ``{Blockchain in internet of things:
  Challenges and Solutions},'' \emph{arXiv:1608.05187 [cs]}, 2016. [Online].
  Available:
  \url{http://arxiv.org/abs/1608.05187{\%}5Cnhttp://www.arxiv.org/pdf/1608.05187.pdf}
\BIBentrySTDinterwordspacing

\bibitem{iota}
``{IOTA},'' \url{https://www.iota.org/get-started/what-is-iota}.

\bibitem{tangle}
A.~Baliga, I.~Subhod, P.~Kamat, and S.~Chatterjee, ``{Performance Evaluation of
  the Quorum Blockchain Platform}.''

\bibitem{cia}
N.~Komninos, E.~Philippou, and A.~Pitsillides, ``Survey in smart grid and smart
  home security: Issues, challenges and countermeasures,'' \emph{IEEE
  Communications Surveys \& Tutorials}, vol.~16, no.~4, pp. 1933--1954, 2014.

\bibitem{quoram}
A.~Baliga, I.~Subhod, P.~Kamat, and S.~Chatterjee, ``{Performance Evaluation of
  the Quorum Blockchain Platform}.''

\bibitem{ethereum}
G.~Wood, ``Ethereum: A secure decentralised generalised transaction ledger,''
  \emph{Ethereum project yellow paper}, vol. 151, pp. 1--32, 2014.

\end{thebibliography}

\end{document}